\documentclass[10pt,aps,pre,twocolumn,superscriptaddress,showpacs,showkeys,floatfix]{revtex4-1}
\usepackage[utf8]{inputenc}
\usepackage{amsmath}
\usepackage{amsfonts}
\usepackage{amssymb}
\usepackage{graphicx}
\usepackage{url}

\begin{document}
  \title{Optical steering of mutual capacitance in nematic liquid crystal cell}
  
  \author{Filip A. Sala}
  \affiliation{Warsaw University of Technology, Faculty of Physics, 00-662 Warsaw, Poland}  
  \author{Marzena M. Sala-Tefelska}
    \affiliation{Warsaw University of Technology, Faculty of Physics, 00-662 Warsaw, Poland}  

\begin{abstract}
In this study a variable capacitor made of Nematic Liquid Crystal cell is proposed and analyzed theoretically. The mutual capacitance is steered with an optical beam of a Gaussian shape launched into the cell. The optical field changes the orientation of the molecules and affects the capacitance. By using Frank-Oseen elastic theory the molecular reorientation is simulated. The influence of various parameters on capacitance such beam width, anchoring condition, externally applied voltage, beam power and launching position is presented. For instance, the maximum tuning range is achieved for wide beams and the molecules initially aligned close to the propagation axis. It is also proved that launching position, especially for narrow beams, has limited influence on capacitance. The proposed component can be used, for instance, in optical power meters, as a feedback in laser or diode systems or just as a variable capacitor in optoelectronic circuits. One of the advantages of this device is that the beam passes through the element, so steering of the capacitance or measuring the parameters of the beam can be realized without splitting the beam. Moreover, due to the low thickness of the liquid crystal layer the attenuation is very low.  
\end{abstract}

\keywords{mutual capacitance, optical steering, liquid crystals}

\pacs{42.70.Df, 84.32.Tt, 07.50.Qx}
\maketitle
\section{Introduction}
Variable capacitors are widely used in electronics, for instance for tuning the resonant circuits or to transform some physical quantities into electric signals. Typically the variable capacitor is tuned mechanically or by applying DC voltage like in capacitance diodes, called varicaps. In this article a design of a new, variable capacitor tuned with optical field is proposed. The setup consists of a cell filled with Nematic Liquid Crystal (NLC). Such materials are anisotropic, typically uniaxial, with two electric permittivities: perpendicular and parallel to the average molecular orientation. It means that the capacitance of such cell depends on the molecular orientation. The actual capacitance of the NLC cell has been already analyzed \cite{Yao2006}. 
\begin{figure}[h]
\begin{center}
\includegraphics[width=0.47\textwidth]{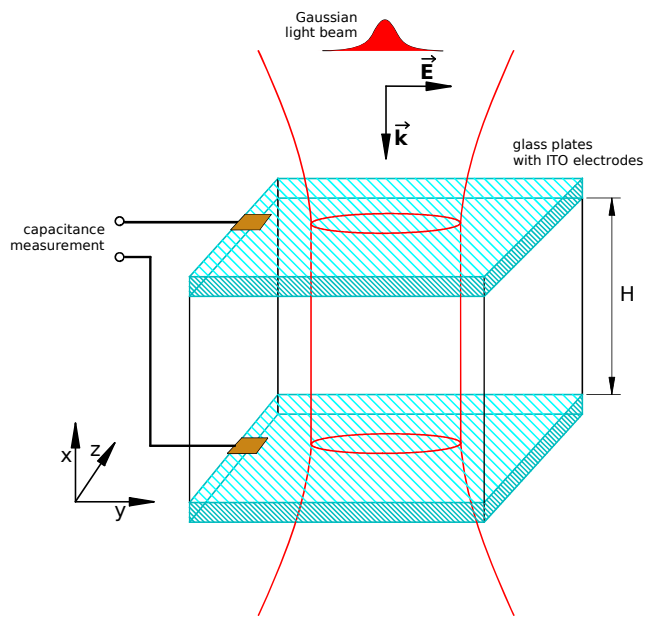}
\caption{Sketch of the analyzed setup: liquid crystal cell of a height $H$ with ITO electrodes.} 
\label{fig:nlc_cell}
\end{center}
\end{figure}
Moreover, NLCs are nonlinear materials, so when the external electric field is applied the molecules change the orientation due to the induced torque. Such mechanism of capacitance steering was also analyzed \cite{Yao2006, Yeh2005, Raynes1979} even with periodic electrodes \cite{Chen1999}. Similar setup can be potentially used for chemical and biological sensors \cite{LC_Sensors2008}. The NLCs not only interact with the electric field, but also with the magnetic field, which was also used for capacitance steering  \cite{MagneticField2013,Shtrikman1971}. The NLC cells were also used for designing thermal diodes \cite{thermal_diode_Melo2016}, NOR and XOR logic gates \cite{Assanto_gates} or multiplexing devices \cite{sala_jnopm2014,ArXiv_spatial_routing_2017}. 

In this article steering of mutual capacitance is achieved by the optical beam of a Gaussian shape launched into the cell. The electric field of the beam interacts with the molecules and changes their orientation. In effect, the capacitance of the whole cell changes. The proposed setup is analyzed theoretically. To model the molecular reorientation the elastic theory is used. 
The influence of the width of the beam, anchoring condition, launching position, beam power and additional external voltage, on the capacitance is presented. The optimal parameters for achieving desired or maximum tuning range and for obtaining response for lower power beams are also described. 

\section{The analyzed setup and governing equations}
\begin{figure}[b]
\begin{center}
\includegraphics[width=.45\textwidth ]{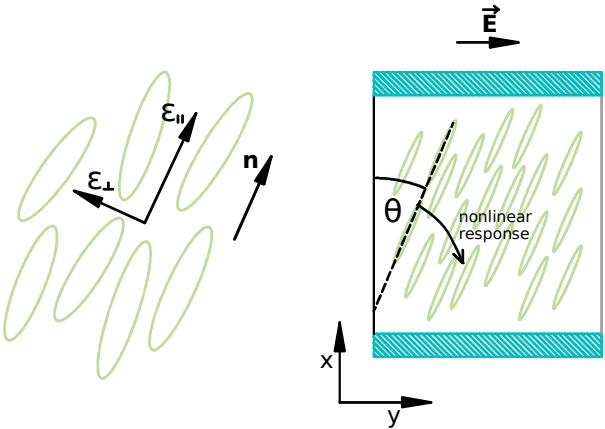}
\caption{Sketch of molecular orientation where $\vec{n}$ - director, and $\varepsilon_\parallel$, $\varepsilon_\perp$ - parallel and perpendicular electric permittivities respectively.}
\label{fig:lc_molecules2}
\end{center}
\end{figure}

\begin{figure}
\centering
\includegraphics[width=0.47\textwidth]{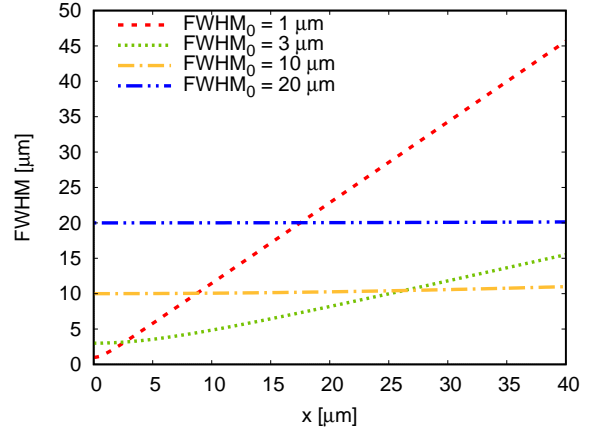}
\caption{Beam width in a linear medium with refractive index $n=1.67$, for wavelength $\lambda=1.55\, \mu \text{m}$ for various $\text{FWHM}_0$ at waist: $1\, \mu \text{m}$ (dashed line), $3\, \mu \text{m}$ (dotted line), $10\, \mu \text{m}$ (dot dashed line) and $20\, \mu \text{m}$ (double dot dashed line). Propagation along the $x$ axis.}
\label{fig:Rayleigh}
\end{figure}

\begin{figure}
\centering
\includegraphics[width=0.47\textwidth]{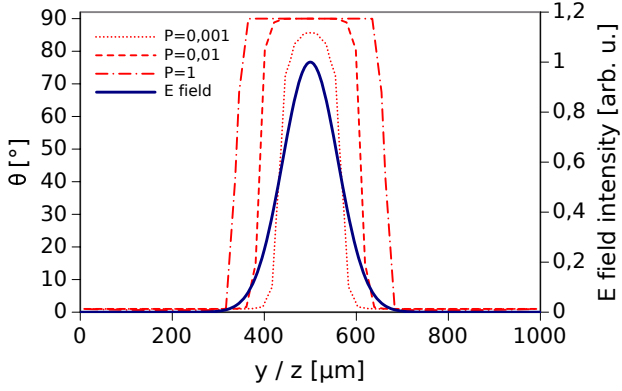}
\caption{Comparison of electric field distribution (solid line) and molecular orientation along y or z axis. Simulations performed for beam width $\text{FWHM}=100\; \mu \text{m}$ and for various input beam powers: $P= 1$ (dot dashed line), $P=0.01$ (dashed line) and $P=0.001$ (dotted line). }
\label{fig:reorientation}
\end{figure}

The theoretically analyzed setup corresponds to the cell, made of glass, of a height $H=5\,\mu \text{m}$ and width and length of $1000\,\mu \text{m}$, filled with a nematic liquid crystal (see Fig.\ \ref{fig:nlc_cell}). On the opposite glass plates the ITO electrodes are deposited, so the cell can be used as a capacitor and its capacitance can also be measured. The average molecular orientation is described by the unit vector, called director, defined as:
\begin{equation}
\vec{n} = [\cos \theta, \sin\theta \sin \phi, \sin \theta \cos\phi] 
\end{equation}
There are strong anchoring conditions on all surfaces of the cell: $\phi_0=\pi/2$ and $\theta_0$. 
Nematic liquid crystals are typically uniaxial and have two electric permittivities: parallel and perpendicular to the director (Fig.\ \ref{fig:lc_molecules2}). The permittivity depends on the frequency, so it is different for high i.e.\ optical frequencies ($\varepsilon_\perp^H$ and $\varepsilon_\parallel^H$) and for low i.e.\ electric frequencies ($\varepsilon_\perp^L$ and $\varepsilon_\parallel^L$).
The light beam of a Gaussian shape is launched into the system along the x axis. In the simulations the width of the beam is assumed constant. In a linear regime it is reasonable, as the analyzed beams are broad with FWHM from $20\,\mu\text{m}$ upto $800\,\mu\text{m}$ . Because the height of the cell is low, about $5\,\mu \text{m}$, even for narrower beams the width is nearly constant along the direction of propagation (see Fig.\ \ref{fig:Rayleigh}). In nonlinear regime, the molecular reorientation affects the refractive index profile forming a waveguide-like distribution. It can lead to a formation of the nondiffracting beams i.e.\ solitons in NLCs called nematicons \cite{AssantoPR,AssantoWiley,tim2007, karpierz2002_physrevE,bennonlocal}. In our case the anchoring conditions at the sidewalls are far from the propagating beam which leads to low nonlocality of molecular reorientation. In Fig.\ \ref{fig:reorientation} comparison of electric field and molecular orientation distributions is shown. Thus, the width of the propagating beam will not be much greater than the width of the input beam. In case of a soliton formation, which is unlikely on a very short distance, the beam will be narrower than the input beam but still of nearly constant width. The so called "soliton breathing" effect, when the width of the beam changes periodically with propagation, is also unlikely due to the short propagation distance. Considering the above the assumption of constant beam width seems valid. 

\begin{figure}
\centering
\includegraphics[width=0.47\textwidth, height=150px]{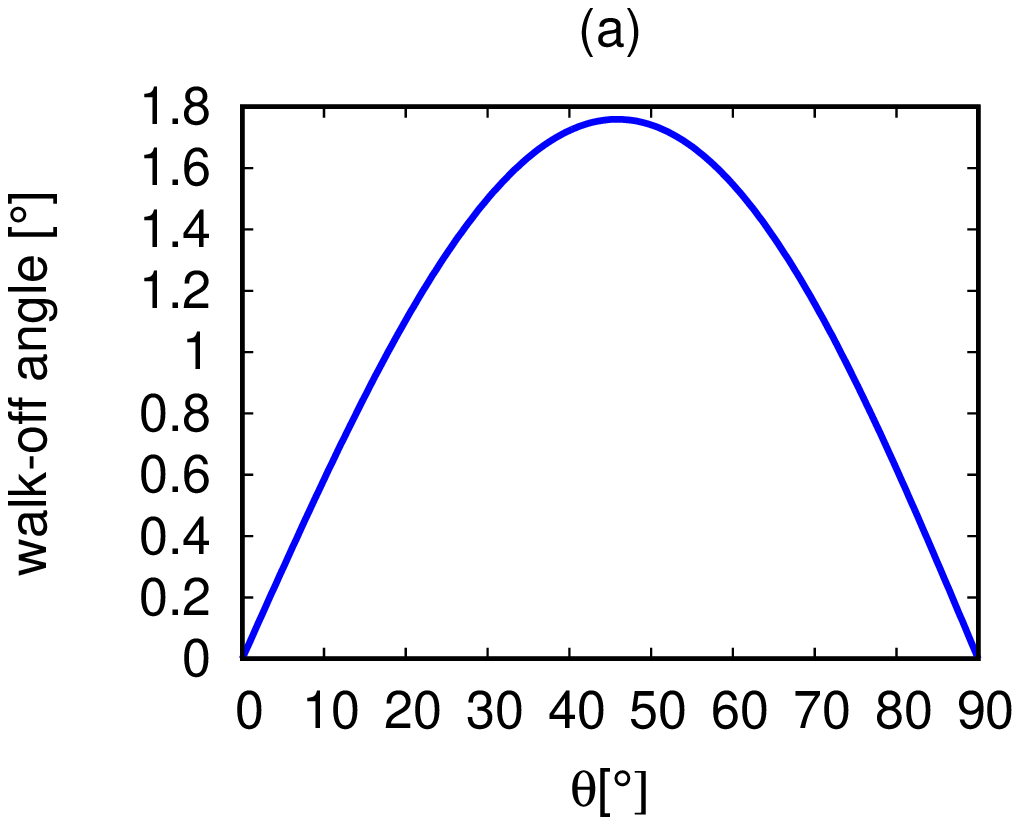}
\includegraphics[width=0.47\textwidth ,height=150px]{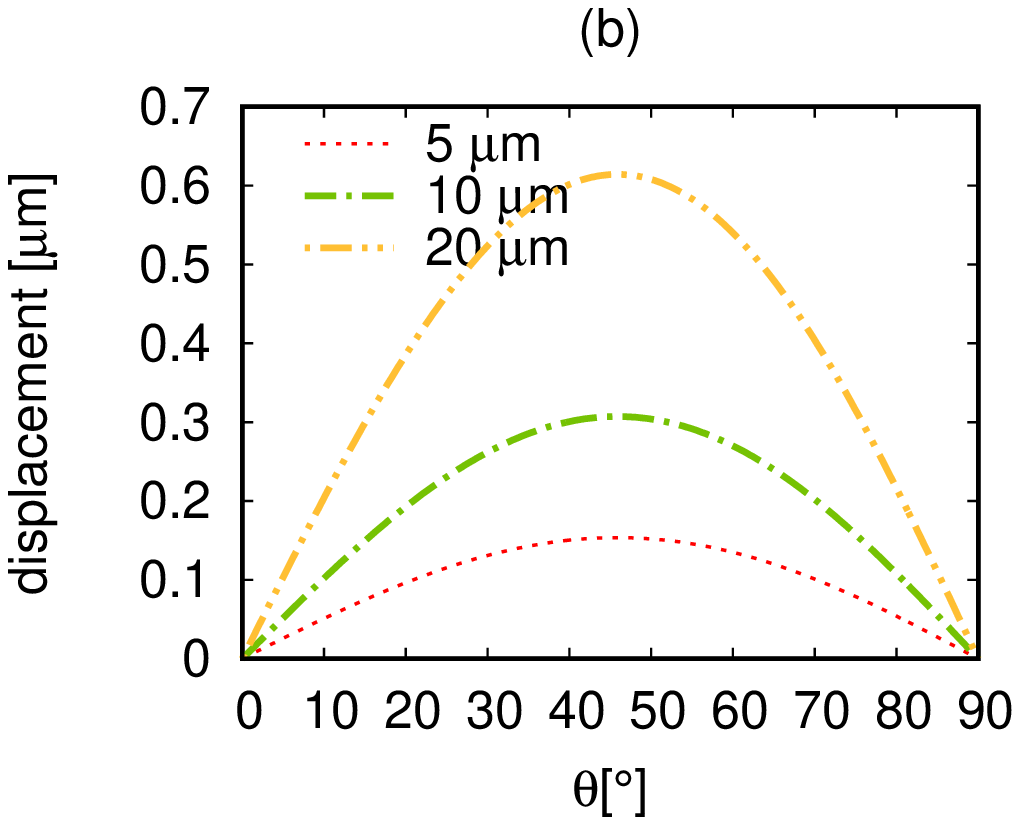}
\caption{(a) Walk-off angle (solid line) calculated by equation (\ref{eq:walkoff}) for various molecular orientations.  (b) Beam displacement at the output of the system for a few cell heights: $H=5\; \mu \text{m}$ (dotted line), $H=10\; \mu \text{m}$ (dot dashed line), $H=20\; \mu \text{m}$ (double dot dashed line).} 
\label{fig:walkoff}
\end{figure}

In anisotropic material when the wavevector $\vec{k}$ and the director $\vec{n}$ are not parallel nor perpendicular to each other, the walk-off phenomenon appears. When light enters into the anisotropic material the wavevector conserves direction but the energy travels at some angle $\delta$ to the wavevector. The walk-off angle can be defined as:
\begin{equation}
\delta = \arctan \left(\dfrac{\Delta \varepsilon \sin 2 \theta}{\Delta \varepsilon +2 \varepsilon_\perp + \Delta \varepsilon \cos2\theta}\right)
\label{eq:walkoff}
\end{equation}
where $\Delta \varepsilon =\varepsilon_\parallel^H - \varepsilon_\perp^{H}$ is optical anisotropy. In Fig.\ \ref{fig:walkoff}a walk-off angle is plotted for various molecular orientations. In Fig.\ \ref{fig:walkoff}b the spatial shift at the output of the system corresponding to the walk-off angle is shown. For the cells even thicker than the analyzed one, the displacement is lower than $1\,\mu \text{m}$, which is negligible corresponding to the transverse dimensions of the cell. Thus, the walk-off phenomenon can also be neglected in the analyses.   

To model the molecular reorientation casued by the optical beam and externally applied voltage the elastic theory is used. The Frank-Oseen equation defining free energy density describes deformation of a liquid crystal \cite{khoo, oseen, frank,  deGennes} 
\begin{eqnarray}
\nonumber f&=&\frac{1}{2}K_{11}(\nabla \vec{n})^2+\frac{1}{2}K_{22}(\vec{n}\cdot (\nabla \times \vec{n}))^2\\\nonumber & &+\frac{1}{2}K_{33}(\vec{n}\times (\nabla \times \vec{n}))^2-\frac{1}{2}\Delta \varepsilon \varepsilon_0 (\vec{n}\cdot \vec{E})^2\\ & &-\dfrac{1}{2}\Delta \varepsilon^L \varepsilon_0(\vec{n} \cdot \vec{E_v})^2 \label{eq:Frank}
\end{eqnarray}
where $K_{11}$,$K_{22}$, $K_{33}$ - are Frank elastic constants corresponding to \textit{splay, twist} and \textit{bend} deformations respectively and $\vec{E}$ is the electric field of an optical beam, $\vec{E_v}=(E_v^x,0,0)$ is the electric field of the externally applied voltage, $\Delta \varepsilon =\varepsilon_\parallel^H - \varepsilon_\perp^{H}$ - optical anisotropy i.e.\ for high frequencies, $\Delta \varepsilon^L= \varepsilon_\parallel^L - \varepsilon_\perp^{L}$ - anisotropy for low frequencies. 
By using the Euler-Lagrange equations (\ref{eq:EulerLagrange}) the extremum of the energy can be found. 
\begin{eqnarray}
\frac{\partial}{\partial x} \frac{\partial f}{\partial \phi_x}+\frac{\partial}{\partial y} \frac{\partial f}{\partial \phi_y}+\frac{\partial}{\partial z} \frac{\partial f}{\partial \phi_z}-\frac{\partial f}{\partial \phi}=0 \nonumber\\
\frac{\partial}{\partial x} \frac{\partial f}{\partial \theta_x}+\frac{\partial}{\partial y} \frac{\partial f}{\partial \theta_y}+\frac{\partial}{\partial z} \frac{\partial f}{\partial \theta_z}-\frac{\partial f}{\partial \theta}=0
\label{eq:EulerLagrange}
\end{eqnarray}
where:
\begin{equation*}
\phi_j=\dfrac{\partial \phi}{\partial j},\quad \theta_j=\dfrac{\partial \theta}{\partial j},\quad j \in \{x,y,z\}  
\end{equation*}
Additionally, assuming that all elastic constants are equal $K_{11}=K_{22}=K_{33}=K$ leads to the following equations \cite{sala_OptExpress2012,sala2012_JOSAB}:
\begin{eqnarray}
\nonumber & &\nabla^2\phi\sin^2 \theta + \sin 2\theta \cdot \left(\dfrac{\partial \phi}{\partial x}\dfrac{\partial \theta}{\partial x}+\dfrac{\partial \phi}{\partial y}\dfrac{\partial \theta}{\partial y}+\dfrac{\partial \phi}{\partial z}\dfrac{\partial \theta}{\partial z}  \right)\\\nonumber & &
+\:\dfrac{\Delta \varepsilon \varepsilon_0}{2K}\bigg[2E_y E_z \sin^2 \theta \cos 2\phi \\\nonumber & &+\: \sin 2\theta (E_x E_y \cos \phi - E_x E_z \sin \phi)  \\& &+\:\sin^2 \theta \sin 2 \phi (E_y^2 - E_z^2)\bigg]=0 \label{eq:phi}\\
& &\nonumber \nabla^2 \theta -\dfrac{1}{2} \sin 2 \theta \left[\left(\frac{\partial \phi}{\partial x}\right)^2 + \left(\frac{\partial \phi}{\partial y}\right)^2 +\left(\frac{\partial \phi}{\partial z}\right)^2 \right] \\\nonumber & &
+\:\dfrac{\Delta \varepsilon \varepsilon_0}{2K}\bigg[E_y E_z \sin 2\theta \sin 2\phi \\\nonumber & &+\: 2\cos 2\theta (E_x E_y \sin \phi +E_x E_z \cos \phi) \\\nonumber & &+\: \sin 2\theta (E_z^2 \cos^2 \phi +E_y^2 \sin^2 \phi - E_x^2 \bigg]\\ & & -\dfrac{\Delta \varepsilon^L \varepsilon_0}{2K} (E_v^x)^2\sin 2\theta = 0
\label{eq:theta}
\end{eqnarray}
The electric field $E$ is complex so $E_jE_k=\Re(E_jE_k^*)=\Re(E_j)\Re(E_k)+\Im(E_j)\Im(E_k)$ where $j,k \in \{x,y,z\}$. The electric field arising  from the external voltage is approximated as constant inside the cell $E_v^x=V_0/H$, where $V_0$ - applied voltage.\\ 
To solve the equations (\ref{eq:phi}) and (\ref{eq:theta}) finite differences in conjunction with SOR (Successive OverRelaxation) method are used \cite{SOR_book}.
The electric field changes the molecular orientation and the electric permittivity tensor, which leads to a change in the capacitance of the cell. To calculate the capacitance, the whole cell is treated as a set of layers of parallelly connected capacitors. The capacitance of such a single layer can be expressed as:
\begin{equation*}
C_{p}(x) =\sum_{yz} \dfrac{\varepsilon_0\varepsilon_\perp^{L} \varepsilon_\parallel^{L}}{\varepsilon_\parallel^{L} \sin^2 \theta(x,y,z)+\varepsilon_\perp^{L} \cos^2\theta(x,y,z)}\dfrac{\Delta y \Delta z}{\Delta x}
\end{equation*}
where $\varepsilon_\parallel^L$, $\varepsilon_\perp^L$ - electric permittivity for low frequencies. The total capacitance of the cell is equal to:
\begin{equation}
C_{tot}=1 \bigg/ \sum_{x} \dfrac{1}{C_p(x)}
\end{equation}

\section{Numerical results}
\begin{figure}
\begin{center}
\includegraphics[width=0.47\textwidth]{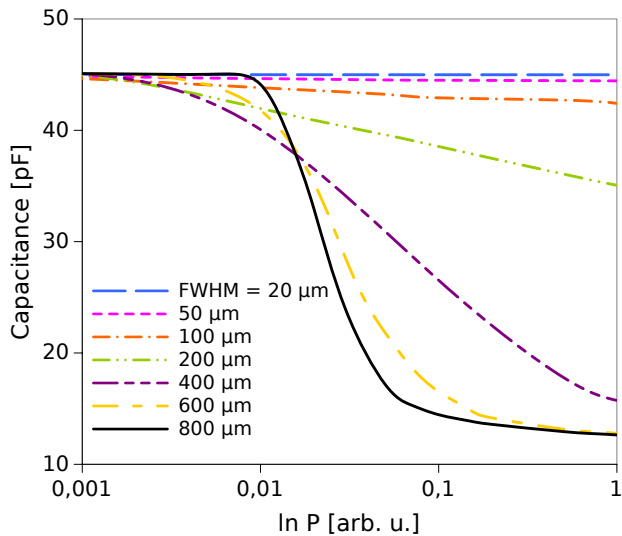}
\caption{The influence of the beam power (logarithmic scale) on the capacitance of the cell for various beam widths (FWHM) and anchoring condition $\theta_0=1^\circ$.}
\label{fig:cap_vs_power}
\end{center}
\end{figure}

\begin{figure}
\centering
\includegraphics[width=0.47\textwidth, height=7cm]{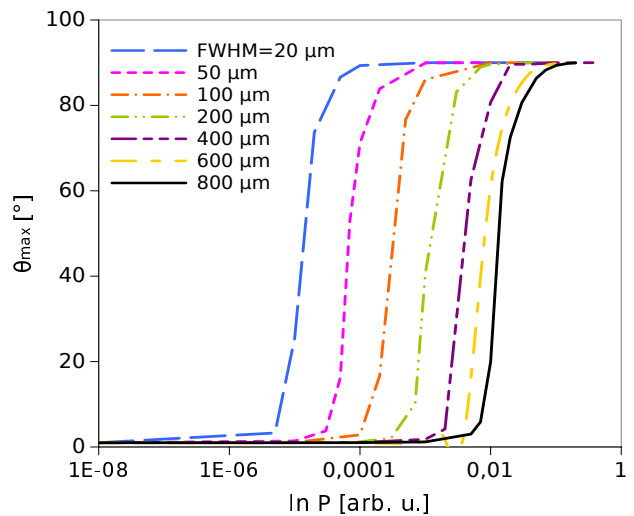}
\caption{The angle of maximum molecular orientation $\theta_{max}$ versus input beam power $P$ in a logarithmic scale. Simulations performed for various beam widths (FWHM) and anchoring condition $\theta_0=1^\circ$.}
\label{fig:thetamax_vs_power}
\end{figure}

The numerical simulations where performed for a liquid crystal with Frank elastic constant $K=5.5\;\text{pN}$ and electric permittivities $\varepsilon_\parallel^H=2.784$, $\varepsilon_\perp^H= 2.294$, $\varepsilon_\parallel^L=25$, $\varepsilon_\perp^L= 5$. About 2500 upto 6000 iterations of the algorithm were sufficient to obtain convergence in most cases. The SOR parameter was set to $\Omega=0.7$. However, only the molecular orientation described by angle $\theta$ affects the capacitance, the reorientation in $yz$ plane was also calculated to obtain more accurate results. In fact, even if the molecules lie in the same plane as the electric field there is always some residual reorientation in perpendicular plane, in this case in $yz$ plane. At first, only the influence of the optical beam, with no external voltage applied $V_0=0$, is analyzed. The beam is launched into the center of the cell. In Fig.\ \ref{fig:cap_vs_power} the cell capacitance is plotted versus input beam power in logarithmic scale. The anchoring condition is $\theta_0=1^\circ$, so the molecules are initially aligned nearly along the $x$ axis. The electric field of the optical beam causes the molecules to reorient and changes the electric permittivity tensor for both optical and low frequencies, in effect of which the capacitance decreases. For narrow beams $\text{FWHM} \leq 50\, \mu \text{m}$ the change of capacitance is very low, even for very high powers. For wider beams $\geq 100\, \mu \text{m}$ the capacitance change is noticeable, upto about $32\,\text{pF}$ for $\text{FWHM} = 800\,\mu\text{m}$.
In Fig.\ \ref{fig:thetamax_vs_power} the angle of maximum molecular orientation in function of beam power is plotted. Although, the change of capacitance for narrow beams is low, the noticeable reorientation appears for lower power than for wide beams. It is caused by the higher power density of narrow beams. It is also clearly visible that reorientation saturates at $\theta=90^\circ$ which leads to saturation of minimum available capacitance (compare with Fig.\ \ref{fig:cap_vs_power}).

\begin{figure}
\centering
\includegraphics[width=0.47\textwidth]{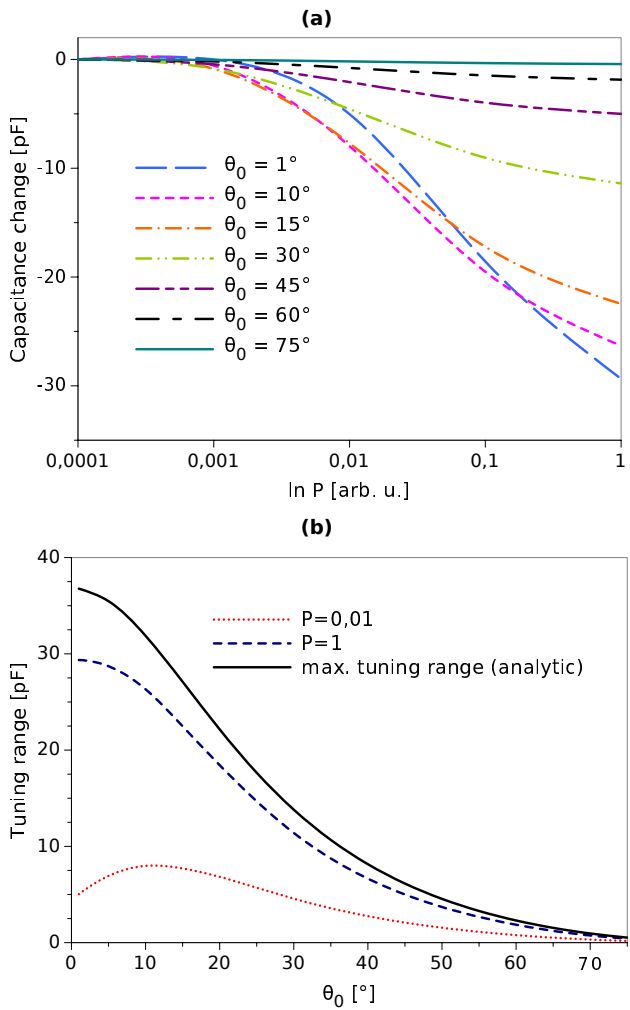}
\caption{(a) Capacitance change with respect to the initial capacitance of the LC cell as a function of input beam power in logarithmic scale. Different curves correspond to different anchoring conditions $\theta_0$, for beam width $\text{FWHM}=400\,\mu \text{m}$. (b) Tuning range i.e.\ difference between maximum and minimum capacitances versus anchoring condition $\theta_0$, for beam width $\text{FWHM}=400\,\mu \text{m}$ and powers: $P=1$ (dashed line) and $P=0.01$ (dotted line). The analytic solution of maximum possible tuning range is plotted with a solid line.} 
\label{fig:cap_vs_power_theta}
\end{figure}

In Fig.\ \ref{fig:cap_vs_power_theta}a the capacitance change versus input beam power is plotted for various anchoring conditions. The maximum capacitance tuning range is plotted in Fig.\ \ref{fig:cap_vs_power_theta}b. According to these results the highest possible tuning range can be obtained for low values of $\theta_0$ i.e.\ for the molecules anchored close to the direction of the $x$ axis. On the other hand, for lower power, for instance $P=0.01$ the maximum tuning range is around $\theta_0=10^\circ$. It means that the maximum tuning range depends on power and for lower power values appears for higher values of anchoring condition $\theta_0$. In such case the tunning range is significantly lower than for higher powers. To obtain significant change in capacitance, the anchoring condition $\theta_0$, in this case, should be lower than about $30^\circ$. To obtain the highest capacitance change the molecules should be aligned nearly along the $x$. In such case the power needed to obtain significant molecular reorientation increases. When the molecules are aligned along the $x$ axis ($\theta_0=0$) the molecular reorientation occurs above Fr\'eedericksz threshold \cite{Leger1972_walls_in_nematics,Freedericksz1927}, which will additionally increase the power needed for reorientation.

\begin{figure}
\centering
\includegraphics[width=0.47\textwidth]{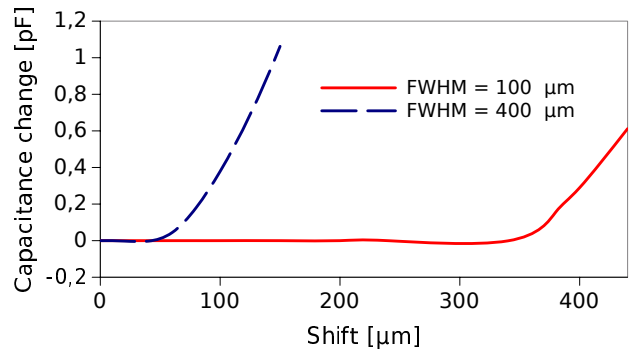}
\caption{The influence of the spatial shift of the beam on the capacitance. Simulations performed for beam widths: $\text{FWHM}=100\,\mu \text{m}$ (solid line) and $\text{FWHM}=400\,\mu \text{m}$ (dashed line) and power $P=0.1$ and $\theta_0=1^\circ$.}
\label{fig:shift}
\end{figure}

The influence of the spatial shift of the beam on capacitance is also analyzed. The results in Fig.\ \ref{fig:shift} show that there is a range of shift values for which the capacitance changes only slightly. The width of this range strongly depends on the width of the beam: it is wider for narrow beams and narrower for wide beams. It means that as long as the beam is launched far from the edge of the cell the capacitance nearly does not depend on the position of the beam.  When the beam is launched closer to the boundary the molecular reorientation is weaker and the total capacitance increases, and the tuning range decreases. There are two mechanisms responsible for these effects: first of all it is much more difficult to reorient molecules near the strong boundary condition and the second, weaker mechanism, is that the tail of the Gaussian beam passes the boundary of the cell, so the total power of the beam, interacting with the liquid crystal, is lower.
\begin{figure}
\centering
\includegraphics[width=0.47\textwidth]{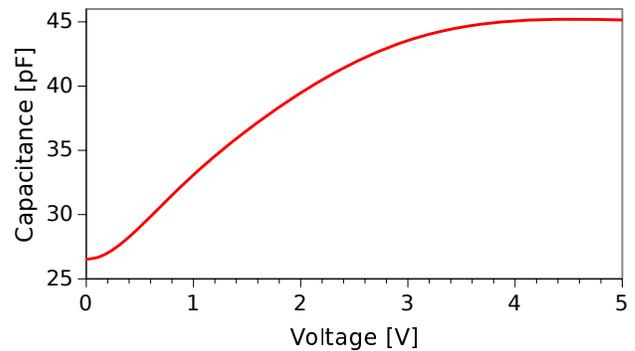}
\caption{Capacitance versus externally applied voltage $V_0$. Simulations performed for beam width $\text{FWHM}=400\,\mu \text{m}$, power $P=0.1$ and $\theta_0=1^\circ$.}
\label{fig:voltage}
\end{figure}

Typically, when using a variable capacitor in an electronic circuit, or even to measure the capacitance, it is needed to apply the external voltage. In a liquid crystal cell the external voltage applied to the ITO electrodes will additionally influence the molecular orientation. This effect was also analyzed and is shown in Fig.\ \ref{fig:voltage}. For very low voltages $V_0<0.2\,\text{V}$ the change in capacitance is negligible but for higher voltages it increases and then saturates to the maximum capacitance of the cell. The NLC cell should rather be used in low voltage applications only, as for higher voltages power needed to obtain the same tuning range is significantly increased. Moreover, the voltages should be in a narrow range, because otherwise the applied signal will strongly influence the capacitance. 
\section{Conclusions}
In the article the analyses of variable capacitor made of NLC cell were presented. It was shown that walk-off phenomenon does not significantly influence the trajectory of the beam and the capacitance of the cell. The beam width was assumed constant, which seems to be valid according to the presented results. To model the molecular reorientation, taking into account the optical and low frequency fields, the elastic theory was used. It was shown that for narrow beams the change in capacitance occurs for lower power values but the maximum tuning range is limited. Much higher tuning range appeared for wide beams, upto $32\,\text{pN}$ for $\text{FWHM}= 800\,\mu \text{m}$. The anchoring condition also plays a crucial role. For high power values the maximum tuning range was achieved for low values of $\theta_0$, so when the molecules were aligned nearly along the $x$ axis. For lower power values the maximum tuning range appeared for higher values of $\theta_0$. In other word, to achieve the highest possible tuning range the molecules should be aligned close to the $x$ axis, but to achieve noticeable change in capacitance, for low powers, higher values of anchoring condition $\theta_0$ have to be chosen. The influence of the launching position of the beam was also analyzed and it was proved that as long as the beam is inputted far from the boundary of the cell, the position does not influence the capacitance. For the beams launched close to the boundary the capacitance increased and the tuning range for the same power ranges decreased. According to our analyses, the external voltage applied to the cell significantly changes capacitance and decreases the tuning range, so the proposed component should be used only for low voltage applications, for the analyzed parameters $V_0<0.2\,\text{V}$. The voltage of applied signals should also be kept in a narrow range to avoid changes of capacitance. The authors are aware that for higher power values the thermal effects will influence the stability, molecular orientation and also the capacitance. To avoid this, materials of possibly the highest electric anisotropy for both optical and low frequencies and low absorption should be used or developed. 

Such device can be applied, for instance, in the optical power meters, as a feedback in laser or diode systems or just as a variable capacitor in optoelectronic circuits. As the beam passes trough the cell, the steering of capacitance or measuring the power or the width of the beam can be realized in-system without splitting the beam. Moreover, low thickness of the cell implies low attenuation. Summarizing, the presented results provide a concept of a variable capacitor steered with an optical beam. The article shows optimal parameters for particular case and methods of calculating them, which can be used in future designs.

\bibliographystyle{ieeetr}
\bibliography{bibliography1}

\begin{thebibliography}{10}

\bibitem{Yao2006}
K.-Y. Lo, C.-C. Shiah, and C.-Y. Huang, ``Actual capacitance function of
  nematic liquid crystal cell,'' {\em Jpn. J. Appl. Phys.}, vol.~45, no.~2R,
  p.~891, 2006.

\bibitem{Yeh2005}
J.~A. Yeh, C.~A. Chang, C.-C. Cheng, J.-Y. Huang, and S.~S.~H. Hsu, ``Microwave
  characteristics of liquid-crystal tunable capacitors,'' {\em IEEE Electron
  Device Lett.}, vol.~26, pp.~451--453, July 2005.

\bibitem{Raynes1979}
E.~P. Raynes, R.~J.~A. Tough, and K.~A. Davies, ``Voltage dependence of the
  capacitance of a twisted nematic liquid crystal layer,'' {\em Mol. Cryst.
  Liq. Cryst.}, vol.~56, no.~2, pp.~63--68, 1979.

\bibitem{Chen1999}
C.-J. Chen, K.~R. Sarma, and A.~Kolosovskaya, ``Capacitance–voltage
  characteristics of liquid crystal displays with periodic interdigital
  electrodes,'' {\em Appl. Phys. Lett.}, vol.~74, no.~1, pp.~147--149, 1999.

\bibitem{LC_Sensors2008}
A.~S. Abu-Abed and R.~G. Lindquist, ``Capacitive transduction for liquid
  crystal based sensors, part ii: Partially disordered system,'' {\em IEEE
  Sens. J.}, vol.~8, pp.~1557--1564, Sept 2008.

\bibitem{MagneticField2013}
N.~Toma\ifmmode \check{s}\else \v{s}\fi{}ovi\ifmmode~\check{c}\else
  \v{c}\fi{}ov\'a, M.~Timko, Z.~Mitr\'oov\'a, M.~Konerack\'a,
  M.~Raj\ifmmode~\check{n}\else \v{n}\fi{}ak, N.~\'Eber, T.~T\'oth-Katona,
  X.~Chaud, J.~Jadzyn, and P.~Kop\ifmmode~\check{c}\else \v{c}\fi{}ansk\'y,
  ``Capacitance changes in ferronematic liquid crystals induced by low magnetic
  fields,'' {\em Phys. Rev. E}, vol.~87, p.~014501, Jan 2013.

\bibitem{Shtrikman1971}
S.~Shtrikman, E.~Wohlfarth, and Y.~Wand, ``Magnetic field dependence of the
  capacity of a twisted nematic liquid crystal cell,'' {\em Phys. Lett. A},
  vol.~37, no.~5, pp.~369 -- 370, 1971.

\bibitem{thermal_diode_Melo2016}
D.~Melo, I.~Fernandes, F.~Moraes, S.~Fumeron, and E.~Pereira, ``Thermal diode
  made by nematic liquid crystal,'' {\em Phys. Lett. A}, vol.~380, no.~38,
  pp.~3121 -- 3127, 2016.

\bibitem{Assanto_gates}
A.~Piccardi, A.~Alberucci, U.~Bortolozzo, S.~Residori, and G.~Assanto,
  ``Soliton gating and switching in liquid crystal light valve,'' {\em Appl.
  Phys. Lett.}, vol.~96, no.~7, p.~071104, 2010.

\bibitem{sala_jnopm2014}
F.~A. Sala, M.~A. Karpierz, and G.~Assanto, ``Spatial routing with
  light-induced waveguides in uniaxial nematic liquid crystals,'' {\em J.
  Nonlinear Opt. Phys. Mater.}, vol.~23, no.~04, p.~1450047, 2014.

\bibitem{ArXiv_spatial_routing_2017}
F.~A. Sala, N.~F. Smyth, U.~A. Laudyn, A.~A.~M. Mirosław A.~Karpierz, and
  G.~Assanto, ``Reorientational solitons in nematic liquid crystals with
  modulated alignment,'' {\em (submitted to Phys. Rev. A.)
  https://arxiv.org/abs/1707.03777}, 2017.

\bibitem{AssantoPR}
M.~Peccianti and G.~Assanto, ``{Nematicons},'' {\em Phys. Rep.}, vol.~516,
  pp.~147--208, 2012.

\bibitem{AssantoWiley}
G.~Assanto, {\em Nematicons, Spatial Optical Solitons in Nematic Liquid
  Crystals}.
\newblock John Wiley and Sons, New York, 2012.

\bibitem{tim2007}
A.~Minzoni, N.~Smyth, and A.~Worthy, ``{Modulation solutions for nematicon
  propagation in non-local liquid crystals},'' {\em J. Opt. Soc. Amer. B},
  vol.~24, pp.~1549--1556, 2007.

\bibitem{karpierz2002_physrevE}
M.~A. Karpierz, ``Solitary waves in liquid crystalline waveguides,'' {\em Phys.
  Rev. E}, vol.~66, p.~036603, 2002.

\bibitem{bennonlocal}
B.~Skuse and N.~Smyth, ``{Interaction of two colour solitary waves in a liquid
  crystal in the nonlocal regime},'' {\em Phys. Rev. A.}, vol.~79, p.~063806,
  2009.

\bibitem{khoo}
I.-C. Khoo, {\em Liquid Crystals: Physical Properties and Nonlinear Optical
  Phenomena}.
\newblock Wiley, New York, 1995.

\bibitem{oseen}
C.~W. Oseen, ``{The theory of liquid crystals},'' {\em Trans. Faraday Soc.},
  vol.~29, pp.~883--899, 1933.

\bibitem{frank}
F.~C. Frank, ``{I. Liquid crystals. On the theory of liquid crystals},'' {\em
  Discuss. Faraday Soc.}, vol.~25, pp.~19--28, 1958.

\bibitem{deGennes}
P.~de~Gennes and J.~Prost, {\em The physics of liquid crystals}.
\newblock Clarendon Press - Oxford, 1993.

\bibitem{sala_OptExpress2012}
F.~Sala and M.~A. Karpierz, ``Modeling of molecular reorientation and beam
  propagation in chiral and non-chiral nematic liquid crystals.,'' {\em Opt.
  Express}, vol.~20, pp.~13923--13938, 2012.

\bibitem{sala2012_JOSAB}
F.~A. Sala and M.~A. Karpierz, ``Chiral and non-chiral nematic liquid crystal
  reorientation induced by inhomogeneous electric fields,'' {\em J. Opt. Soc.
  Am. B}, vol.~29, no.~6, pp.~1465--1472, 2012.

\bibitem{SOR_book}
L.~Hageman and D.~Young, {\em Applied Iterative Methods}.
\newblock New York: Academic Press, 1981.

\bibitem{Leger1972_walls_in_nematics}
L.~Leger, ``Static and dynamic behaviour of walls in nematics above a
  {F}reedericks transition,'' {\em Solid State Commun.}, vol.~11,
  pp.~1499--1501, 1972.

\bibitem{Freedericksz1927}
V.~Fr{\'e}edericksz and A.~Repiewa, ``Theoretisches und experimentelles zur
  frage nach der natur der anisotropen fl{\"u}ssigkeiten,'' {\em Zeitschrift
  f{\"u}r Physik}, vol.~42, pp.~532--546, Jul 1927.

\end{thebibliography}

\end{document}